\documentclass[aps,pra,reprint,superscriptaddress,a4paper]{revtex4-1}

\usepackage{graphicx,xcolor}
\usepackage{hyperref}

\newcommand{\ket}[1]{|#1\rangle}

\newcommand{\POLIMI}{Dipartimento di Fisica - Politecnico di Milano, p.za Leonardo da Vinci 32, 20133 Milano, Italy}
\newcommand{\IFN}{Istituto di Fotonica e Nanotecnologie - Consiglio Nazionale delle Ricerche (IFN-CNR), p.za Leonardo da Vinci 32, 20133 Milano, Italy}
\newcommand{\SAPIENZA}{Dipartimento di Fisica - Sapienza Universit\`{a} di Roma, p.le Aldo Moro 5, 00185 Roma, Italy}
\newcommand{\contreq}{These authors contributed equally.}

\begin{document}

\title{Integrated sources of entangled photons at telecom wavelength\\ in femtosecond-laser-written circuits}

\author{Simone Atzeni}
\altaffiliation{\contreq}
\affiliation{\POLIMI}
\affiliation{\IFN}
\author{Adil S.Rab}
\altaffiliation{\contreq}
\affiliation{\SAPIENZA}
\author{Giacomo Corrielli}
\affiliation{\IFN}
\affiliation{\POLIMI}
\author{Emanuele Polino}
\affiliation{\SAPIENZA}
\author{Mauro Valeri}
\affiliation{\SAPIENZA}
\author{Paolo Mataloni}
\affiliation{\SAPIENZA}
\affiliation{\IFN}
\author{Nicol\`{o} Spagnolo}
\affiliation{\SAPIENZA}
\author{Andrea Crespi}
\affiliation{\POLIMI}
\affiliation{\IFN}
\author{Fabio Sciarrino}
\email{fabio.sciarrino@uniroma1.it}
\affiliation{\SAPIENZA}
\author{Roberto Osellame}
\email{roberto.osellame@polimi.it}
\affiliation{\IFN}
\affiliation{\POLIMI}

\begin{abstract}
Photon entanglement is an important state of light that is at the basis of many protocols in photonic quantum technologies, from quantum computing, to simulation and sensing. The capability to  generate entangled photons in integrated waveguide sources is particularly advantageous due to the enhanced stability and more efficient light-crystal interaction. Here we realize an integrated optical source of entangled degenerate photons at telecom wavelength, based on the hybrid interfacing of photonic circuits in different materials, all inscribed by femtosecond laser pulses. We show that our source, based on spontaneous parametric down-conversion, gives access to different classes of output states, allowing to switch from path-entangled to polarization-entangled states with net visibilities above 0.92 for all selected combinations of integrated devices.
\end{abstract}

\maketitle

\section{Introduction}

Entanglement is a powerful feature of quantum systems and a key resource in quantum information science being widely exploited to perform exclusive quantum protocols as teleportation \cite{teleportation1}, entanglement swapping \cite{swapping1} and repeaters or to outperform classical performances in metrology \cite{metrology1}, criptography \cite{crypto1} and computation \cite{comp1,comp2}. In the last decade, the unique properties of entanglement have stimulated the development of efficient and versatile sources of entanglement carriers \cite{source1,source2,source3}. In this framework, photons represent a favourable choice due to weak decoherence, the possibility of implementing entanglement in several degrees of freedom and the availability of efficient and robust optical components for routing and manipulation of the generated photons. In particular, huge advances have been achieved in integrated optics allowing to realize complex linear circuits with dozens of components on the same chip. This evolution paved the way to the realization of quantum optics experiments, as boson sampling \cite{BS1,BS2,BS3,BS4,BS5,BS6,BS7,BS8,BS9} and quantum random walks \cite{QRW1,QRW2,QRW3}, otherwise not achievable with bulk optics approach. To further scale up the complexity and fully capitalize on advantages of the integrated optics approach, quantum photonics is moving towards the integration of sources on chip as well.
%Adding sources on chip represents a promising strategy to capitalize on advantages of integrated optics approach in quantum photonics.
In bulk optics, sources of entangled photon pairs can be naturally achieved by spontaneous parametric down conversion (SPDC) in nonlinear crystals \cite{kwiatentangledsource}; a promising strategy to realize their integrated counterparts is represented by employing waveguides in nonlinear crystals. Due to the enhanced light-matter interaction, a boost in source brightness adds up to the standard advantages of the integrated approach, as miniaturization and optical phase stability. Although it has been demonstrated that path-entangled states can be generated by down conversion in waveguide sources \cite{coupled1,coupled2}, generation of polarization entanglement in an integrated device requires additional effort. In fact, the integrated sources proposed in the literature, generate polarization entanglement either outside the chip \cite{kaiser,Herrman} or based on non-degenerate photon pairs \cite{Martin}. Recently, an on-chip source of degenerate entangled photons has been demonstrated in lithium niobate waveguides, however the pumping scheme is not yet fully integrated requiring a bulk Sagnac loop configuration \cite{sansonippln}.

Here, we report on the fabrication and characterization of a fully integrated optical source for path- and polarization-entangled photon-pairs. Exploiting femtosecond-laser-written photonic circuits, we demonstrate the flexibility of the interferometric approach \cite{jinsource, meanysource,panossource} to generate different quantum states of light and the feasibility of a hybrid-material assembly to develop high-performance and re-arrangeable microsystems for quantum information science.

\begin{figure*}
\centering
\includegraphics[width=0.99\textwidth]{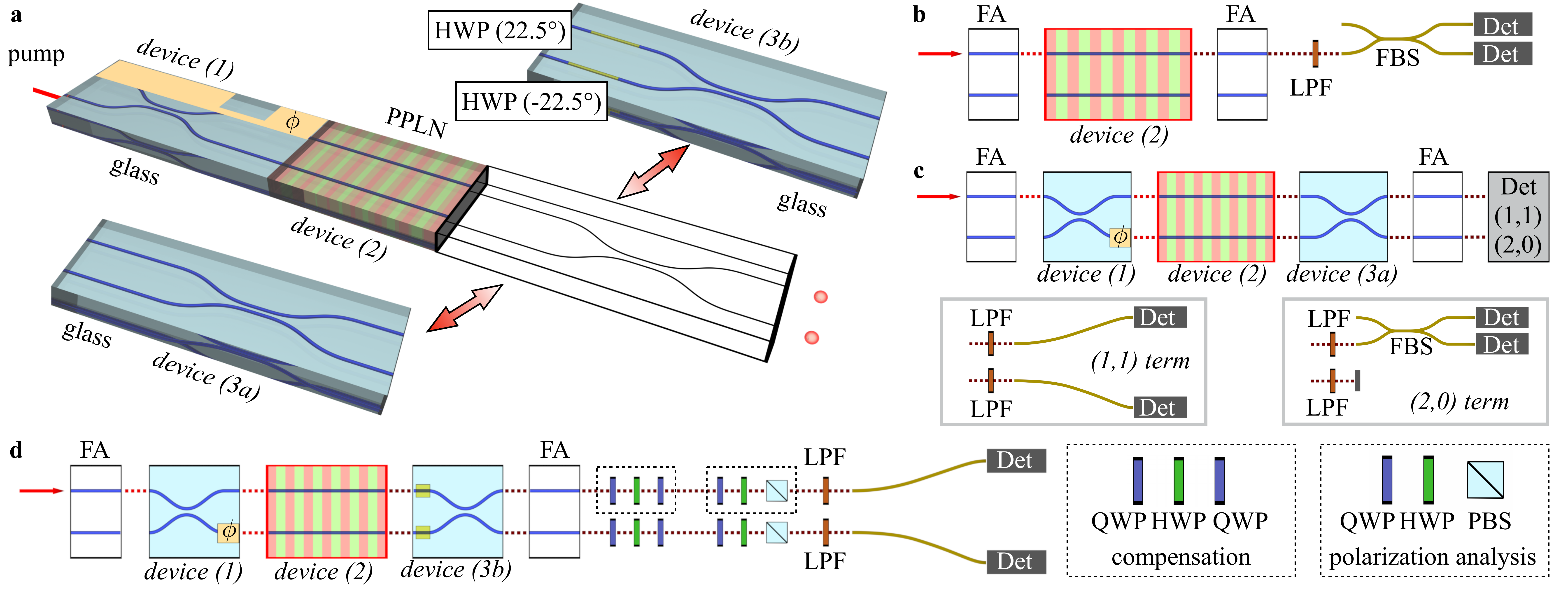}
\caption{{\bf a}, Overall scheme of the integrated source, comprising three cascaded integrated devices. Device (3a) or (3b) can be exchanged depending on the desired output state. {\bf b}, Apparatus employed to characterize the photon pair generation in PPLN waveguides. Device (2) is directly interfaced with input and output FAs. The output of each waveguide is directly sent to the detection apparatus, where a FBS separates the output to two detectors to discriminate two-photon events. {\bf c}, Apparatus for the characterization of the output state by inserting device (3a). The output state, coupled via input and output FAs, is detected to alternatively characterize the (1,1) and (2,0)/(0,2) terms, the latter by inserting a FBS on the measured mode. {\bf d}, Apparatus for the characterization of the polarization-entangled-state generated when device (3b) is used. The output state, collected by a FA, undergoes polarization compensation through a set of wave-plates, and is then analyzed in polarization by means of wave-plates and a PBSs. Legend: PPLN - periodically poled lithium niobate; FA - fiber array; LPF - long pass filter; HWP - half-wave plate; QWP - quarter-wave plate; PBS - polarizating beam-splitter; FBS - fiber beam-splitter; Det - detector.}
\label{Fig1}
\end{figure*}

\section{Layout of the source}

The single photon generation system is composed of three integrated devices (Fig. \ref{Fig1}a): (1) reconfigurable balanced directional coupler at 780~nm; (2) two identical waveguides in a periodically poled lithium niobate (PPLN) chip; (3a-b) a third interchangeable device operating at 1560~nm for the preparation of different output states. The combination of such devices permits the generation of single photon states at telecom wavelength and state engineering through a reconfigurable Mach-Zehnder interferometer. Details on the fabrication of these devices are reported in Appendix A.

The first directional coupler splits the pump equally to feed the two laser-written waveguides in the PPLN second device. Single-photon pairs are generated in both waveguides through a Type 0 SPDC process. Dynamical control of the phase between the two paths is ensured by a thermal shifter fabricated in the first device. The third device closes the interferometer and recombines the generated photons to obtain the desired output. We employed two different devices, giving access to different classes of output states. In a first case, the third device (3a) consists in a balanced directional coupler, leading to an output state of the form:
\begin{equation} \label{path-state}
\ket{\psi_{(3a)}} = \frac{\ket{0,2} - \ket{2,0}}{\sqrt{2}}\cos({\frac{\phi}{2}) + \ket{1,1} \sin(\frac{\phi}{2})},
\end{equation}
where $\vert i,j \rangle$ stands for a state with $i$ and $j$ photons on the two waveguides respectively. Here, a NOON state or a product state $\ket{1,1}$ can be selected controlling the phase $\phi$ between the two output arms of the balanced directional coupler in the first device. 

In the latter case, the third device (3b) is composed by an half-wave plate at $22.5^{\circ}$ (on mode 1), an half-wave plate at $-22.5^{\circ}$ (on mode 2), and a balanced polarization insensitive directional coupler.
Conditioned to the detection of a single-photon on each output mode of the device, the output state is a polarization-entangled state of the form:
\begin{equation}
\ket{\psi_{(3b)}} = \frac{1}{\sqrt{2}} (\vert +,+ \rangle + e^{\imath \phi} \vert -,- \rangle) ,
\end{equation}
where $\vert \pm \rangle = 2^{-1/2} (\vert H \rangle \pm \vert V \rangle)$ are diagonal linear polarization states at $45^{\circ}$. As in the previous case, $\phi$ can be tuned by the reconfigurable phase in the first device.

\section{Characterization of the nonlinear waveguides}

A wavelength-tunable Ti:Sapphire oscillator operating in the continuous-wave regime (CW) is coupled to the integrated device by means of single mode fiber arrays in order to characterize the generation of photon pairs in each PPLN waveguide (Fig. \ref{Fig1}b). Long pass filters are used at the output to remove the residual pump beam from the generated photon pairs, which are then sent to the detection apparatus. The generated pairs are measured by coupling one PPLN waveguide at a time with a fiber beam splitter (FBS). Two fold detection is performed by two avalanche photodiodes: the first operates in free running at efficiency $\eta_{\mathrm{eff}}^1=25\%$ (ID230 by ID Quantique) with a dead time of 10$\mu$s, while the second one is employed in external gating mode at efficiency $\eta_{\mathrm{eff}}^2=25\%$ (ID210 by ID Quantique) with a dead time of 30$\mu$s. This configuration allows to reduce the dark counts rate and to maximize the detection efficiency. Detection of one photon on one mode of the FBS at the output of a PPLN waveguide triggers the second detector, on the other FBS mode, for the detection of the second photon. An appropriate time delay is introduced to allow the communication between the detectors. By controlling the internal trigger delay and the gate width of the second detector it is possible to optimize the signal to noise ratio (SNR). The generation rate of the PPLN waveguides has been verified individually with 30$\mu$W of pump power. The obtained detection rate for each waveguide is $\sim$26000Hz single counts and $\sim$10Hz coincidences, measured after the FBS. Optimizing the detection parameters, we achieved a maximum SNR value of $\sim 140$ for the two fold coincidences.

\section{Generation of path-entangled pairs}

\begin{figure}
\centering
\includegraphics[width=0.49\textwidth]{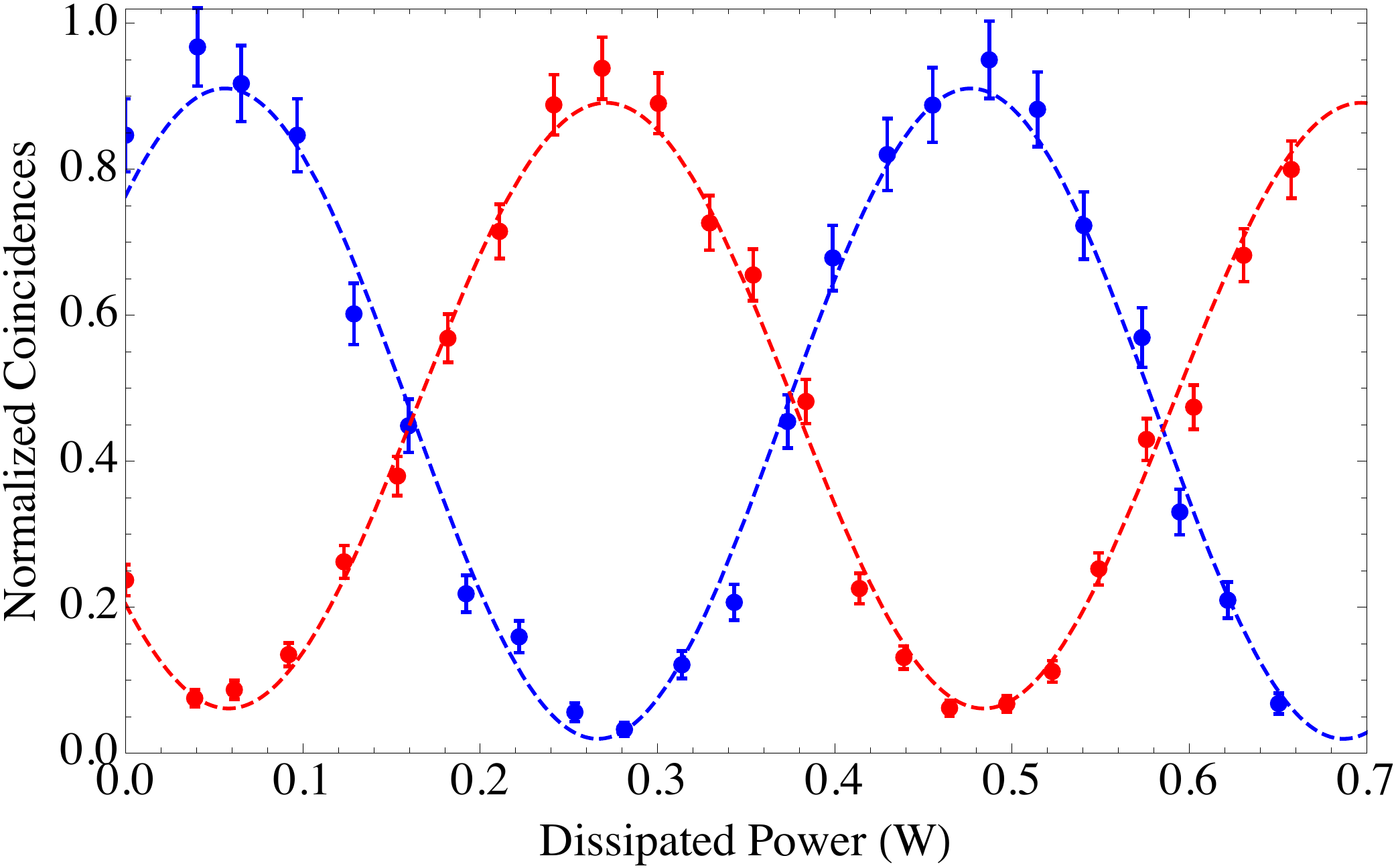}
\caption{Interference fringes in the path-entangled configuration for the states $\ket{1,1}$ (in red) and $\ket{0,2}$ (in blue), as a function of the dissipated power in the actively tunable phase of the first chip. Dots represent experimental raw data with respective error bars, while dashed lines correspond to the fitted curves.}
\label{Fig3}
\end{figure}

We then performed the characterization of the final output state reported in Eq.~(\ref{path-state}). This is achieved by exploiting the configuration of Fig.~\ref{Fig1}c, thus connecting device (3a) after the PPLN waveguide structures. We measured separately the $\ket{1,1}$ and the $\ket{0,2}$ terms as a function of the dissipated power by the thermal resistors. The $\ket{1,1}$ contribution was measured directly at the two outputs of the system, while the $\ket{0,2}$ contribution was measured by coupling one output to a in-fiber beam splitter. We measured the active tuning between $\ket{0,2}$ and $\ket{1,1}$ as a function of the reconfigurable phase $\phi$. The possibility to engineer the path-entangled state is highlighted by the anti-phase oscillations of the two contributions in the coincidence counts, corresponding to $\cos^{2} (\frac{\phi}{2})$ and  $\sin^{2} (\frac{\phi}{2})$, respectively (Fig. \ref{Fig3}). The visibilities of the coincidence oscillations are $\mathcal{V}_{\ket{1,1}}^{\mathrm{raw}}=0.877 \pm 0.004$ and $\mathcal{V}_{\ket{0,2}}^{\mathrm{raw}}=0.935 \pm 0.003$ for raw measurements. Subtracting accidental coincidences, the visibilities become: $\mathcal{V}_{\ket{1,1}}=0.970 \pm 0.004$ and $\mathcal{V}_{\ket{0,2}}=0.980 \pm 0.004$ showing the high quality of the generated state.

\section{Generation of polarization-entangled pairs}

\begin{figure}
\centering
\includegraphics[width=0.49\textwidth]{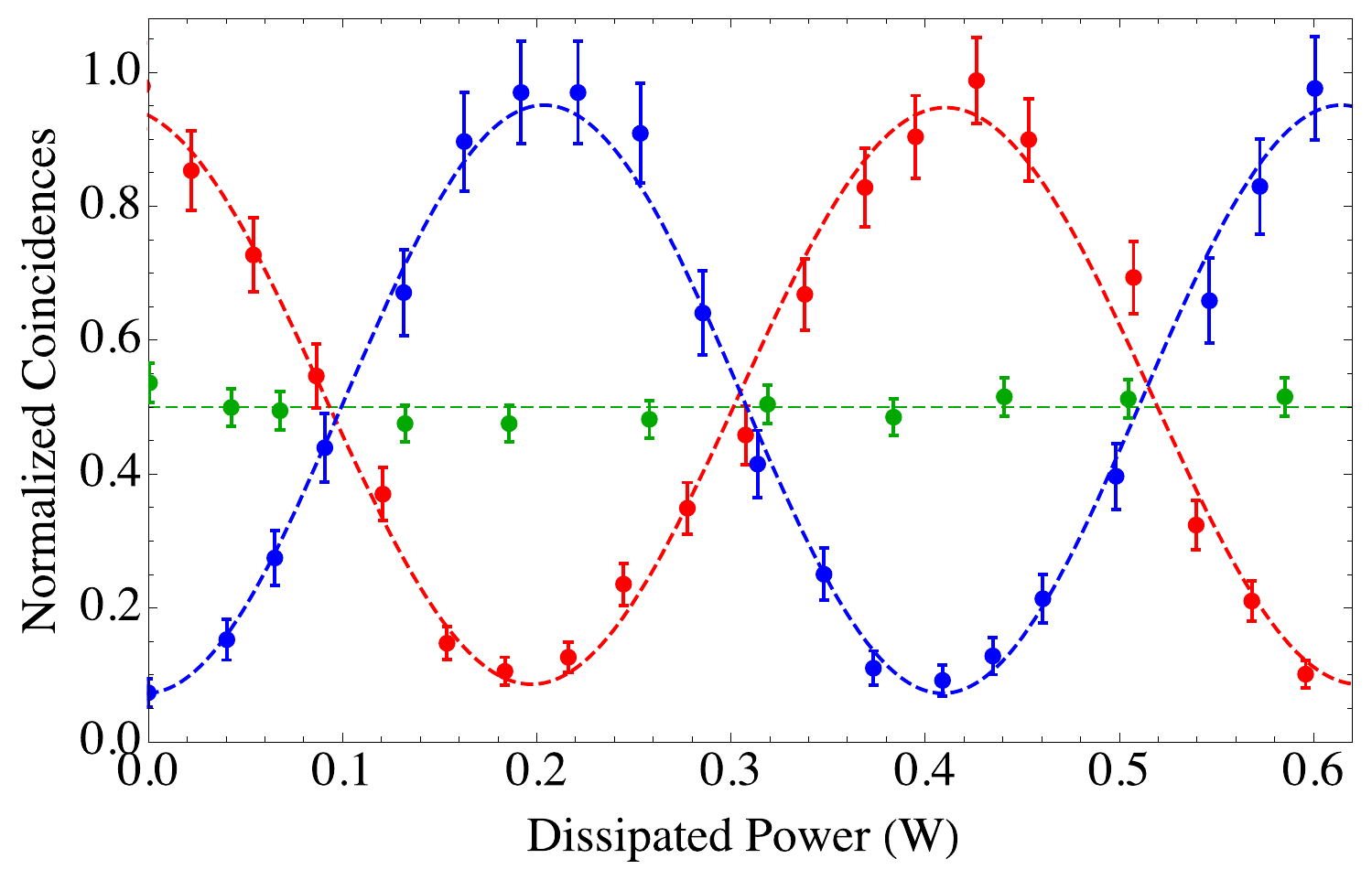}
\caption{Fringe pattern in the polarization-entangled configuration obtained by measuring the contribution $\vert +,+\rangle$ (red) and the contribution $\vert +,- \rangle$ (blue) of the output state, as a function of the dissipated power in the actively tunable phase of the first chip. Fringe pattern obtained by measuring the contribution $\vert H,H\rangle$ (green) of the output state. Dots are raw experimental data, while dashed lines correspond to the fitted curves.}
\label{Fig4}
\end{figure}

We have then performed the characterization of the polarization entangled state, which is obtained conditioned to the detection of a photon on each output mode, according to the setup of Fig. \ref{Fig1}d. In this case, device (3b) is employed after the PPLN waveguides. Before being analyzed and detected, the generated state undergoes polarization compensation to cancel undesired rotations occurring in the output fiber array. During the polarization compensation stage we also rotated the basis of the entangled pair so as to obtain an output state of the form $2^{-1/2} (\vert H,H \rangle + e^{\imath \phi} \vert V,V \rangle)$. For $\phi=0$ ($\phi=\pi$) this corresponds to polarization Bell state $\vert \phi^{+} \rangle$ ($\vert \phi^{-} \rangle$). In order to characterize the generated output state, we first measured the fringe pattern as a function of the tunable phase $\phi$ in two different polarization bases. More specifically, in the diagonal $\vert \pm \rangle$ basis the output state is expected to present a sinusoidal oscillation pattern, while a measurement in the $\vert H/V \rangle$ basis should present no dependence on the phase $\phi$. The experimental results are shown in Fig.\ref{Fig4} and are in agreement with the expected behaviour. The measured visibilities in the different bases are $\mathcal{V}^{\mathrm{raw}}_{\vert +,+ \rangle} = 0.858 \pm 0.019$ and $\mathcal{V}^{\mathrm{raw}}_{\vert +,- \rangle} = 0.834 \pm 0.018$ for raw data ($\mathcal{V}_{\vert +,+ \rangle} = 0.957 \pm 0.015$ and $\mathcal{V}_{\vert +,- \rangle} = 0.929 \pm 0.017$ by subtracting for accidental coincidences). Furthermore, we observe that the pattern in the $\vert H/V \rangle$ basis is almost constant. These results provide evidences of the source correct operation.

To further characterize the generated state, we chose a specific value for the phase $\phi=\pi$, corresponding to the generation of the Bell state $\vert \phi^{-} \rangle$. Hereafter, all experimental value are referred to experimental data subtracting for accidental coincidences. We first measured the expectation values of Pauli matrices products $\langle \sigma_{i} \otimes \sigma_{i} \rangle$, where $i=X,Y,Z$, that correspond to evaluating polarization correlations in three different bases. We obtained $\langle \sigma_{X} \otimes \sigma_{X} \rangle = 0.942 \pm 0.008$ ($\vert H/V \rangle$ basis), $\langle \sigma_{Y} \otimes \sigma_{Y} \rangle = 0.895 \pm 0.010$ ($\vert \pm \rangle$ basis) and $\langle \sigma_{Z} \otimes \sigma_{Z} \rangle = 0.944 \pm 0.008$ ($\vert R/L \rangle$ basis), showing the presence of correlation in all three bases. This allows to apply an entanglement test on the generated state \cite{Eisen2004}, namely $S = \sum_{i=X,Y,Z} \vert \langle \sigma_{i} \otimes \sigma_{i} \rangle \vert \leq 1$ for all separable states. The experimental value is $S_{\mathrm{exp}} = 2.782 \pm 0.015$, thus violating the inequality by $\sim 115$ standard deviations and confirming the presence of polarization entanglement. Furthermore, full state tomography is reported in Appendix B.

%\section*{Results}

\section{Discussion}
In this paper we have proposed unusual Mach-Zehnder interferometers to generate different quantum states of light as product, path- and polarization-entangled photon states. In particular, we have shown the versatility and the modularity of the proposed strategy based on the combination of integrated optical circuits realized in different materials. We then validated the adopted design realizing the photonic chips by femtosecond laser micromachining and characterizing properties of the resulting quantum states. We expect that these results will encourage the choice of hybrid approach in the realization of integrated entangled sources, since as a matter of principle it allows to maximize performances of these devices by choosing the best substrate and component for each application. In perspective, a pumping system could be directly added to the device making it become a ready to use integrated source of spectrally degenerate entangled photons. 
%and paving the way to fully integrated quantum technologies. 

%In principle, this hybrid approach allows to easily enhance the performance by choosing the best substrate and component for each application. 

%Exploiting the capability of femtosecond laser micromachining in processing different transparent materials we realize all the devices and we demonstrate the correct behaviour of the proposed sources. 
%Improving the quality of integrated devices (i.e. reduction of losses,  

\begin{acknowledgments}
We thank Dr. D. Gatti and R. Gotti for their help in the measurement of second harmonic generation process.

This work was supported by the ERC-Advanced Grant CAPABLE (Composite integrated photonic platform by femtosecond laser micromachining; grant agreement no. 742745), by the H2020-FETPROACT-2014 Grant QUCHIP (Quantum Simulation on a Photonic Chip; grant agreement no. 641039, http://www.quchip.eu), and by the Marie Curie Initial Training Network PICQUE (Photonic Integrated Compound Quantum Encoding, grant agreement no. 608062, funding Program: FP7-PEOPLE-2013-ITN, http://www.picque.eu).
\end{acknowledgments}

%\section*{Additional Information}
%\subsection*{Appendix A: 
\appendix
\section{Photonic devices fabrication}
\label{subsec:method}

Photonic circuits were inscribed by direct femtosecond laser micromachining using a Yb:KYW cavity-dumped mode-locked oscillator ($\lambda$~=~1030~nm), that produces ultrafast pulses (300~fs pulse duration, 1~MHz repetition rate). 
In the first device, ultrashort pulses with 220~nJ pulse energy were focused with a microscope objective (0.6~NA, 50$\times$) 25~$\mu$m below the surface of a transparent aluminum-borosilicate glass substrate (Corning, EAGLE XG), allowing the fabrication of single mode waveguides with mode size (1/e$^2$) of 8$\times$9~$\mu$m$^2$ at the fundamental wavelength . Relative translation of the sample with a velocity of 40~mm~$s^{-1}$ (Aerotech FiberGLIDE 3D) permits to design the balanced directional coupler. Tuning pulse energy (370~nJ) and depth (170~$\mu$m), waveguides with single mode behaviour at down-converted wavelength (single gaussian mode of almost circular profile with 15.5~$\mu$m 1/e$^2$ diameter) were realized in the same material allowing the realization of device (3a). Coupling losses have been estimated to be 0.37~dB with respect to single mode fibers with mode dimension of 10.4$\times$10.4 $\mu$m$^2$, while measured propagation losses are 0.3~dB/cm.
Exploiting the same femtosecond laser source and the same setup used for waveguide fabrication, the thermo-optic phase shifter was fabricated according to the method described in Ref.~\cite{flamini2015}. The component presents a resistance of 75 Ohm and it allows to obtain a phase shift of 2$\pi$ with a dissipated power of about 400~mW.

Tilted waveguides were realized underfilling an oil-immersion microscope objective 1.4~NA, 100$\times$ (according to the method described in Ref.~\cite{corrielli2014}). They present a birefringence of about 3~$\times$~10$^{-5}$ and at a suitable length (25~mm) they behave as integrated half-wave plates: $>$98\% of the incoming H polarized light, impinging on the $22.5^{\circ}$ or $-22.5^{\circ}$ waveguide, is converted into D or A polarization state, respectively. A polarization insensitive balanced directional coupler is fabricated in cascade of integrated HWPs following the procedure described by Ref.~\cite{tacca}. The size of device (3b) is 37~mm~$\times$~12~mm and it presents IL of about 4.5~dB.

\begin{figure*}[p]
\centering
\includegraphics[width=0.49\textwidth]{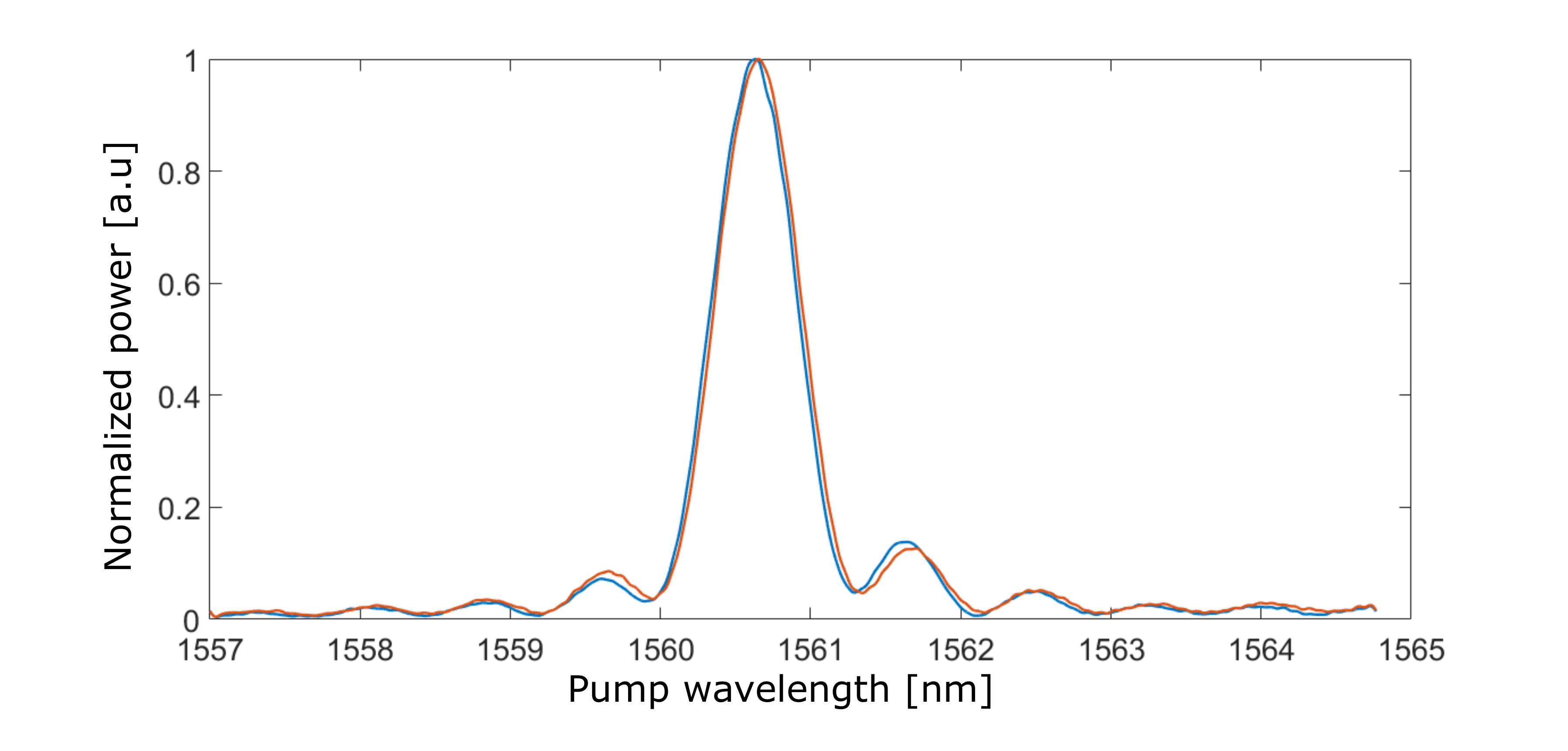}
\caption{Second harmonic spectrum of waveguide 1 (blue) and waveguide 2 (red). The overlap integral between the two spectra is above 99\%, showing very high indistinguishability in nonlinear properties of the pair. The degenerate process will be phase-matched at $\lambda_p$ = 780.31 $\pm$ 0.01 nm $\rightarrow$ 1560.62 $\pm$ 0.02 nm at room temperature.}
\label{FigS1}
\end{figure*}

A pair of waveguides in the active nonlinear material (z-cut periodically-poled magnesium-doped lithium niobate, MSHG1550-1.0-20, Covesion Ltd.) were realized by a multiscan technique as the one described in Ref.~\cite{osellame2007}. Nonlinear waveguides, inscribed in a region with a poling period of $\Lambda$ = 19.5~$\mu$m, present a mode size of 12$\times$12 $\mu$m$^2$ at 780 nm wavelength and 18$\times$16~$\mu$m$^2$ at 1550 nm. Since the peak wavelength of second harmonic generation spectrum is directly related to the degenerate wavelength of spontaneous parametric down conversion, nonlinear behaviour was analyzed through SHG process. Thus, we retrieved the wavelength of the degenerate process, which will be phase-matched at $\lambda_p$ = 780.31~$\pm$ 0.01~nm $\rightarrow$ 1560.62 $\pm$ 0.02~nm at room temperature. The comparison between the SHG curves, that exhibits an overlap above 99~\%, underlines the almost identical nonlinear properties of fabricated waveguides (Fig. \ref{FigS1}). The length of active device is 18~mm.

%\subsection*{Appendix B: Quantum state tomography} 
\section{Quantum state tomography} 
\label{subsec:tom}
\begin{figure*}[p]
\centering
\includegraphics[width=0.49\textwidth]{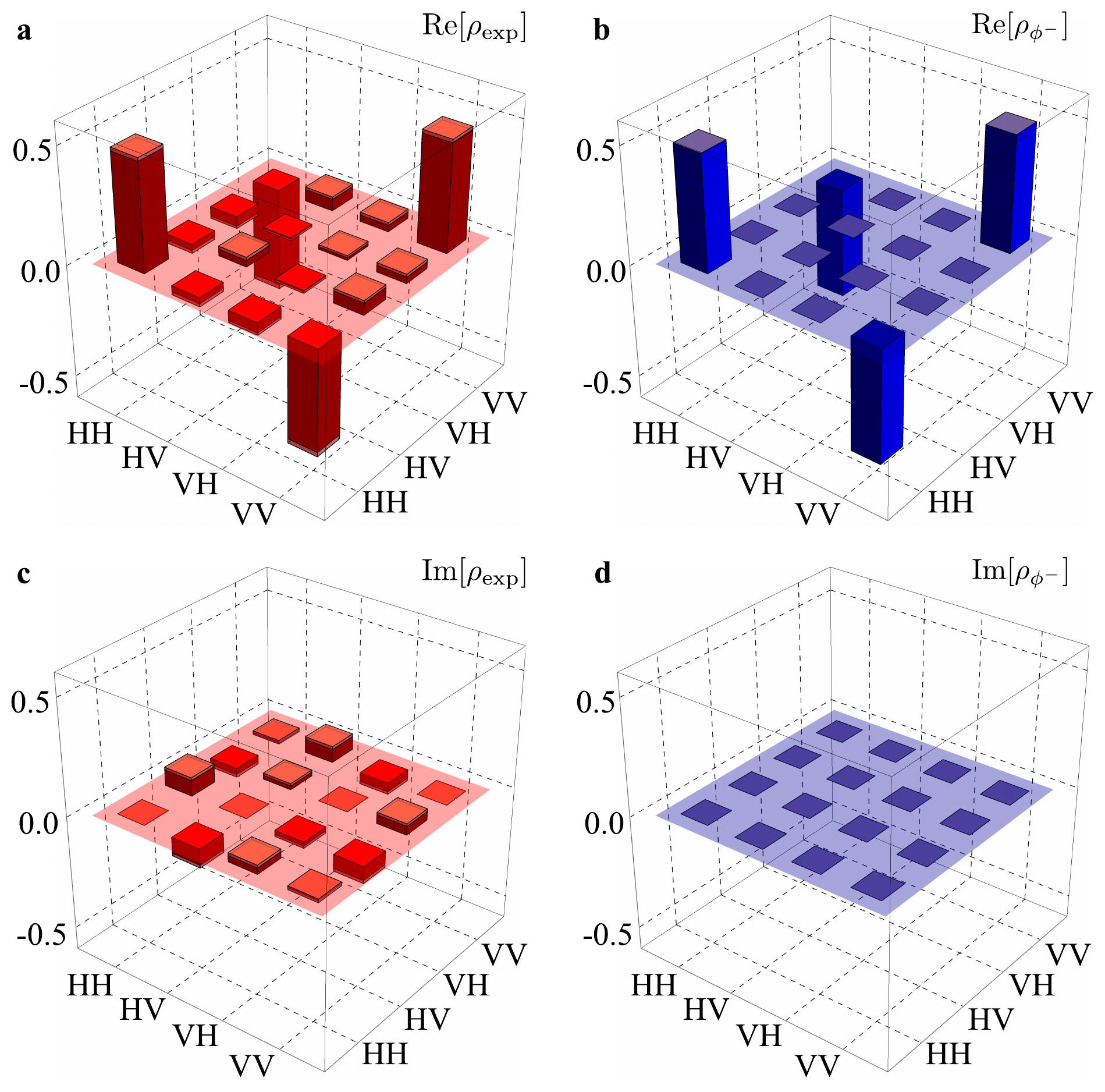}
\caption{\textbf{a-d,} Results of the quantum state tomography performed for a value of the phase set to $\phi=\pi$, corresponding to the $\vert \phi^{-} \rangle$ Bell state. \textbf{a,} Real part of the experimental density matrix $\rho_{\mathrm{exp}}$. \textbf{b,} Real part of the theoretical density matrix $\rho_{\phi^{-}}$.  \textbf{c,} Imaginary part of the experimental density matrix $\rho_{\mathrm{exp}}$. \textbf{d,}. Imaginary part of the theoretical density matrix $\rho_{\phi^{-}}$.}
\label{FigS2}
\end{figure*}

We  performed quantum state tomography \cite{James2001} to fully reconstruct the state density matrix. The results are shown in Fig. \ref{FigS2}a-d, where the obtained density matrix $\rho_{\mathrm{exp}}$ is compared with the one $\rho_{\phi^{-}}$ for an ideal state. We achieved a value of the fidelity $\mathcal{F}(\rho_{\mathrm{exp}},\rho_{\phi^{-}}) = \mathrm{Tr}[(\sqrt{\rho_{\mathrm{exp}}} \rho_{\phi^{-}} \sqrt{\rho_{\mathrm{exp}}})^{1/2}]^{2}$ between theory and experiment equal to $\mathcal{F}=0.929 \pm 0.011$, thus showing the quality of the generated state. The purity and amount of entanglement are quantified respectively by $\Psi=\mathrm{Tr}[\rho_{\mathrm{exp}}^{2}]= 0.908\pm 0.018$ and by the concurrence $C=0.905 \pm 0.022$.

\end{document}